\documentclass[seceq,twocolumn]{jpsj2} 
\def\runauthor{Hiroaki \textsc{Kusunose}}

\title{
Influence of gap structures to specific heat in oriented magnetic fields:\\
Application to the orbital dependent superconductor, Sr$_2$RuO$_4$
}

\author{
Hiroaki \textsc{Kusunose}\thanks{kusu@cmpt.phys.tohoku.ac.jp}
}

\inst{
Department of Physics, Tohoku University, Sendai 980-8578
}

\abst{
We discuss influence of modulation of gap function and anisotropy of Fermi velocity to field angle dependences of upper critical field, $H_{c2}$, and specific heat, $C$, on the basis of the approximate analytic solution in the quasiclassical formalism.
Using 4-fold modulation of the gap function and the Fermi velocity in the single-band model, we demonstrate field and temperature dependence of oscillatory amplitude of $H_{c2}$ and $C$.
We apply the method to the effective two-band model to discuss the gap structure of Sr$_2$RuO$_4$, focusing on recent field angle-resolved experiments.
It is shown that the gap structures with the intermediate magnitude of minima in $[100]$ direction for $\gamma$ band, and tiny minima of gaps in $[110]$ directions for $\alpha$ and $\beta$ bands give consistent behaviors with experiments.
The interplay of the above two gaps also explains the anomalous temperature dependence of in-plane $H_{c2}$ anisotropy, where the opposite contribution from the passive $\alpha\beta$ band is pronounced near $T_c$.
}

\kword{
gap symmetry, oriented magnetic field, quasiclassical theory, multiband, Sr$_2$RuO$_4$
}

\begin{document}
\maketitle

\section{Introduction}
A determination of its gap symmetry has been a long standing issue in a course study of unconventional superconductivity.
Because of a modulation of gap amplitude, thermodynamic properties at low temperature follow power-law behaviors, which give a hint of the structure of the gap function \cite{Leggett75,Sigrist91}.
However, this cannot determine an absolute direction of existing gap minima in principle.
Moreover, it is rather difficult to distinguish practically higher power-law behaviors due to point nodes from exponential behaviors of isotropic gap.
Recently, a powerful method has been developed to measure the modulation of gap amplitude directly.
Namely, field angle dependences in oriented magnetic fields have shown oscillatory behaviors in thermal conductivity \cite{Izawa02a,Izawa02b,Izawa01a,Izawa01b,Tanatar01a,Tanatar01b} and specific heat \cite{Deguchi03,Aoki03,Park03a,Park03b,Park03c,Deguchi04}, depending on detail structure of the gap function.

In order to analyze the oscillatory behaviors, the so-called Doppler shift method has been used frequently, in which the energy shift in the low-energy quasiparticle (QP) spectrum due to supercurrent flowing around vortices is taken into account \cite{Volovik93,Vekhter99,Won01}.
The method neglects the scattering by vortex cores and the overlap of the core states \cite{Dahm02,Miranovic03}.
It makes the theory to be valid in low-temperature and low-field limit, and it still remains unknown what is the range of its applicability in the $(T,H)$ phase diagram.
On the other hand, at fields near $H_{c2}$ the analytic solution for spatially averaged quantities in the quasiclassical formalism is available with the reasonable approximation \cite{Brandt67,Pesch75,Houghton98,Dahm02}.
The present author extended the approximate analytic method to anisotropic singlet and unitary triplet pairings as well as multiband superconductivity \cite{Kusunose04}.
Comparisons with reliable numerical calculations over all field range at the lowest temperature indicated that the method is competent for a semi-quantitative analysis of experiments in wide region of the ($T,H$) phase diagram \cite{Kusunose04}.

The purpose of this paper is to investigate the influence of the gap structure and the anisotropy of the Fermi velocity to the oscillatory behaviors in $H_{c2}$ and $C$ on the basis of the approximate analytic solution.
With those knowledge we discuss the most plausible gap structures of the layered perovskite superconductor, Sr$_2$RuO$_4$, which has been surveyed extensively with use of the rotating magnetic fields \cite{Mackenzie03,Izawa01b,Deguchi03,Deguchi04,Tanatar01a,Tanatar01b,Mao00,Lupien01,Yaguchi02}.

There now exists considerable evidence that Sr$_2$RuO$_4$ pronounces the spin-triplet superconductivity (SC) with broken time reversal symmetry, for instance, the unchanged spin susceptibility in the Knight shift across $T_c=1.5$ K \cite{Ishida98}, the intrinsic magnetization detected by $\mu$SR with the onset of SC \cite{Luke98}, and the absence of a coherence peak in $1/T_1T$ \cite{Ishida97}.

Another important aspect of Sr$_2$RuO$_4$ is the orbital dependent superconductivity (ODS), which comes from three quasi two-dimensional cylindrical Fermi surfaces called $\alpha$, $\beta$ and $\gamma$ \cite{Agterberg97,Zhitomirsky01}.
The latter is almost decoupled from the formers due to different parity of $t_{2g}$ orbitals with respect to the conducting layer.
The ODS scenario can resolve the inconsistency between the power-law temperature dependences compatible with lines of tiny gap, and almost isotropic thermal conductivity under in-plane magnetic fields.
Moreover, the ODS also explains the peculiar $T^2$ dependence of the penetration depth, which is incompatible with lines of tiny gap alone \cite{Kusunose02}.
Recently, Deguchi and co-workers have measured the field-angle dependence of the specific heat to give a direct measure of thermally excited QP in accordance with the SC gap structure \cite{Deguchi03,Deguchi04}.
They suggest that Sr$_2$RuO$_4$ exhibits the ODS and its primary SC gap in $\gamma$ band has minima in [100] direction, while the passive $\alpha$, $\beta$ bands have tiny gaps in [110] direction.

Concerning the angle-resolved measurements, the anomalous temperature dependence of the in-plane $H_{c2}$ anisotropy has been observed where its oscillatory amplitude with maximum in [110] direction ($H_{c2}^{[100]}=1.5$ T) decreases at elevated temperature and even changes sign near $T_c$ \cite{Mao00}.
Although few theoretical arguments have been made for this puzzle in terms of the tetragonal symmetry breaking of the two-dimensional order parameter \cite{Agterberg98,Sigrist00}, a direct evidence of the scenario has not been observed yet.
It is important to recognize that a reasonable model should accompany correct field angle dependences with corresponding $H_{c2}$ anisotropy together with its anomalous temperature dependence.

The paper is organized as follows.
In the next section, we introduce our model of the gap function and the anisotropy of the Fermi velocity, then we give necessary formulas to discuss field angle dependence.
In \S 3 we demonstrate the influence of the gap structure and the anisotropy of the Fermi velocity to the field and the temperature dependence of oscillatory amplitude of $H_{c2}$ and $C$ in the single-band model, keeping the primary $\gamma$ band of Sr$_2$RuO$_4$ in mind.
In \S 4 we discuss the most plausible gap structures for Sr$_2$RuO$_4$ shedding a light on recent field angle-resolved experiments.
The last section summarizes the paper.

\section{Model and formulation}
\begin{figure}[tb]
\includegraphics[width=8.5cm]{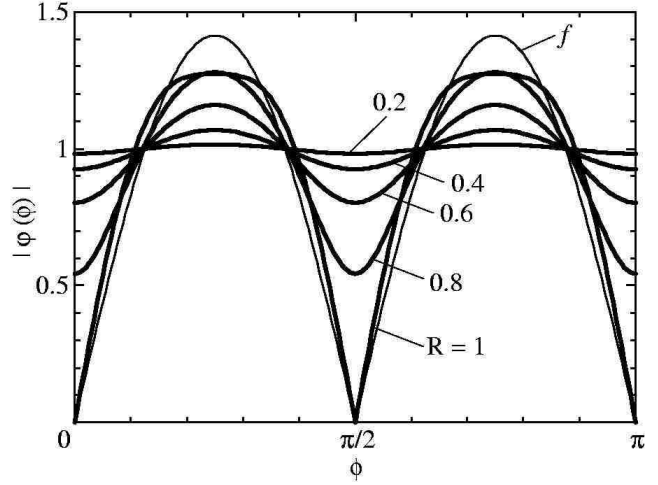}
\caption{The angle dependence of the gap magnitude for various $R$. The thin line represents the case of $f$ wave for comparison.}
\label{gap}
\end{figure}

Let us begin with the model of the gap function. The ${\mib d}$ vector of the triplet pairing is factorized as ${\mib d}(\hat{\mib k})=\hat{\mib z}\Delta \varphi(\hat{\mib k})$, where the angle dependence of the gap function, $\varphi(\hat{\mib k})$ is normalized as $\langle |\varphi(\hat{\mib k})|^2\rangle=1$, $\langle\cdots\rangle$ being the average over the Fermi surface.
Among various type of gap functions proposed so far we adopt the model proposed by Miyake and Narikiyo for $\gamma$ band \cite{Miyake99},
\begin{multline}
\varphi(\hat{\mib k})=\sqrt{\frac{\sin^2[\pi R\cos(\phi-\phi_0)]+\sin^2[\pi R\sin(\phi-\phi_0)]}{1-J_0(2\pi R)}}\\
\times {\rm sgn}[\sin(\phi-\phi_0)],
\label{gap-shape}
\end{multline}
where $J_0(x)$ is the Bessel function of zeroth order, $\phi$ is the azimuthal angle of $\hat{\mib k}$, and $\phi_0$ denotes the position of one of the gap minima (we adopt $\phi_0=0$, i.e., [100] direction for $\gamma$ band).
Although the original meaning of $R$ is the radius of the Fermi circle in unit of $\pi/a$, $a$ being the lattice constant, we regard it as a phenomenological parameter to characterize the magnitude of the gap minima.
Note that the results in the present paper depends only on the magnitude of the gap, $|\varphi(\hat{\mib k})|$, so that there is no difference between isotropic $s$ wave and the chiral $p$-wave for example.
The angle dependence of the gap magnitude is shown in Fig.~\ref{gap}.
$R=1$ corresponds to the gap with line of zeros, while $R=0$ is the isotropic full gap.
For comparison, the dependence of $f$ wave, $\varphi(\hat{\mib k})=\sqrt{2}\sin(2\phi)e^{i\phi}$, is also shown in Fig.~\ref{gap}.
We will also apply the same form of eq.~(\ref{gap-shape}) to the secondary gap in the $\alpha\beta$ bands with the minima in [110] direction, i.e., $\phi_0=\pi/4$.
The microscopic mechanism for these type of gap functions is discussed by Nomura and Yamada using the third-order perturbation theory in the 3-band Hubbard model \cite{Nomura02}.

\begin{figure}[tb]
\includegraphics[width=8.5cm]{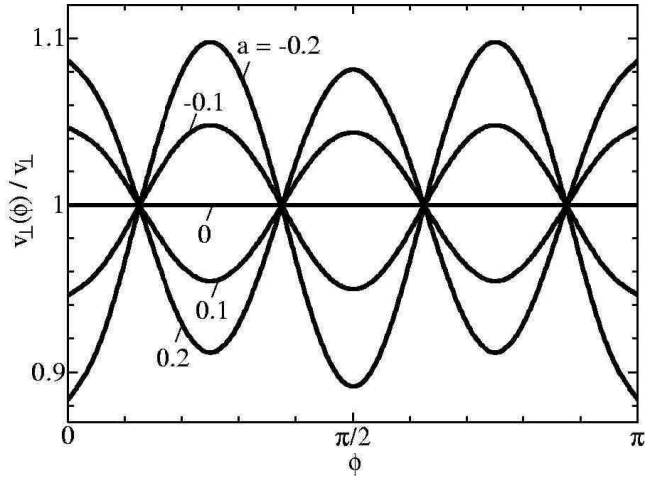}
\caption{The in-plane anisotropy of the Fermi velocity}
\label{v-a}
\end{figure}

Next, we introduce the in-plane anisotropy of the Fermi velocity.
The essential feature of the anisotropy with 4-fold symmetry is described by
\begin{multline}
v_{x,y}(\hat{\mib k})=\sqrt{\frac{2}{2+2a+a^2}}v_{\perp}\times\\
\times \biggl[(1+a\cos2\phi)\cos\phi,(1-a\cos2\phi)\sin\phi \biggr].
\end{multline}
The anisotropy of $v_{\perp}(\phi)=\sqrt{v_x^2+v_y^2}$ is shown in Fig.~\ref{v-a}.
In the case of Sr$_2$RuO$_4$, $a$ is small positive for $\gamma$ band, and almost isotropic, $a\simeq0$, for $\alpha\beta$ bands \cite{Mazin97}.
The modulation of ${\mib v}$ along $z$ direction is approximated roughly by $v_z(\hat{\mib k})=v_z{\rm sgn}(k_z)$ for quasi two-dimensional Fermi surfaces.
The two dimensionality is characterized by the ratio, $\chi=\langle v_z^2\rangle/\langle v_x^2\rangle=2v_z^2/v_\perp^2$.
The average of the Fermi velocity is defined as $v=\sqrt{\langle v_x^2+v_y^2+v_z^2 \rangle}=\sqrt{v_\perp^2+v_z^2}$.

We now explain briefly the approximate analytic solution of the fundamental quasiclassical equations for weak-coupling superconductivity under magnetic fields.
The solution can be obtained by the following approximations: (i) the spatial dependence of the internal magnetic field is averaged by $B$, (ii) the vortex lattice structure is expressed by the Abrikosov lattice, and (iii) the diagonal element of the Green's function is approximated by the spatial average.
Although all approximations mentioned above is valid near $H_{c2}$, comparisons with reliable numerical calculations suggest that the solution is competent quantitatively in wide region of the $(T,H)$ phase diagram except in very low $T$ and $H$ regions \cite{Kusunose04}.
For further detail we recommend the reader to refer the literature \cite{Brandt67,Pesch75,Houghton98,Dahm02,Kusunose04}.
Here we only quote the expression of the free energy measured from the normal state \cite{Kusunose04}, which is given for strongly type-II superconductors, $B\simeq H$, in the clean limit as,
\begin{multline}
\Omega_{\rm SN}/N_0=\sum_{\ell\ell'}(\lambda^{-1})_{\ell\ell'}\Delta_\ell \Delta_{\ell'}
+\sum_\ell n_\ell \biggl[ \Delta_\ell^2 \ln\left(\frac{T}{T_c}\right) \\
-\Delta_\ell^2\ln\left(\frac{2e^\gamma\omega_c}{\pi T_c}\right)+2\pi T \sum_{n=0}^\infty\left(\frac{\Delta_\ell^2}{\omega_n}-\langle I_\ell \rangle \right)
 \biggr],
\label{free}
\end{multline}
where $\ell$ is the band suffix (either $\gamma$ or $\alpha\beta$), $N_0$ is the total density of states (DOS) in the normal state, $n_\ell=N_{0\ell}/N_0$ is the partial DOS of the band $\ell$, and $\omega_n=(2n+1)\pi T$ is the fermionic Matsubara frequency.
Here $\lambda$ is the interaction matrix in unit of $N_0^{-1}$, $\gamma=0.577216$ is the Euler's constant and the cut-off $\omega_c/\pi T_c=100$ is used in the summation of the Matsubara frequency.
The function $I_\ell$ is given by
\begin{equation}
I_\ell=\frac{2g_\ell}{1+g_\ell}\sqrt{\pi}\left(\frac{2\Lambda}{\tilde{v}_{\ell\perp}(\hat{\mib k})}\right)\Delta_\ell^2|\varphi_\ell(\hat{\mib k})|^2W(iu_{\ell n}),
\end{equation}
with
\begin{equation}
g_\ell=\left[
1+\frac{\sqrt{\pi}}{i}\left(\frac{2\Lambda}{\tilde{v}_{\ell\perp}(\hat{\mib k})}\right)^2\Delta_\ell^2|\varphi_\ell(\hat{\mib k})|^2W'(iu_{\ell n}),
\right]^{-1/2}
\end{equation}
where $\Lambda=(2|e|H)^{-1/2}$ is the magnetic length, $u_{\ell n}=2\Lambda \omega_n/\tilde{v}_{\ell\perp}(\hat{\mib k})$, and $W(z)=e^{-z^2}{\rm erfc}(-iz)$ is the Faddeeva function.
Here $\tilde{v}_{\ell\perp}(\hat{\mib k})$ is the component of ${\mib v}$ perpendicular to the field, which is given for the in-plain field, ${\mib H}=H(\cos\phi_h,\sin\phi_h,0)$ as
\begin{multline}
\left( \frac{\tilde{v}_{\ell\perp}(\hat{\mib k})}{v} \right)^2=\frac{2\sqrt{\chi}}{2+\chi}\biggl[1+\frac{4}{2+2a+a^2}\times\\
\times\biggl(\sin(\phi-\phi_h)-a\cos2\phi\sin(\phi+\phi_h)\biggr)^2
\biggr].
\label{vperp}
\end{multline}
Note that without the anisotropy in the velocity, $a=0$, the oscillatory behaviors depend only on the relative angle, $\phi_0-\phi_h$.
In the single-band model the difference of $\chi$ can be absorbed in the definition of $H_{c2}$ because $\tilde{v}_{\perp}$ appears only in the form of $\Lambda/\tilde{v}_\perp$.
This is not the case for multiband models where $\chi_\ell$ is generally different.
Once we obtain the equilibrium free energy for given $T$ and ${\mib H}$ by minimizing the free energy with respect to the gap magnitude $\Delta_\ell$, we can discuss the oscillatory behaviors of $H_{c2}$ and $C$.

\section{Results in single-band model}
Let us discuss the single-band model with the gap having minima in [100] direction ($\phi_0=0$).
At $T=0$ the in-plane $H_{c2}$ anisotropy due to the modulation of the gap and/or the anisotropy of ${\mib v}$ is given by the formula\cite{Kusunose04},
\begin{equation}
H_{c2}=\frac{2}{|e|}\left(\frac{\pi v}{T_c}\right)^2\exp\left[-\left\langle |\varphi(\hat{\mib k})|^2\ln\left(\frac{\tilde{v}_{\perp}(\hat{\mib k})}{v}\right)^2\right\rangle-\gamma\right].
\end{equation}
From this expression the minima of $H_{c2}$ are obtained when the minima of the oscillation in the gap and in the perpendicular component of the velocity coincide.
Thus, without the in-plane anisotropy in ${\mib v}$, i.e., $a=0$, the minima of $H_{c2}$ are simply realized when the field is parallel to the direction of the gap minima.
On the other hand, without the gap modulation, i.e., $R=0$, the anisotropy of ${\mib v}$ gives rise to the minima of $H_{c2}$ when $\phi_h\parallel[110]$ ($\phi_h\parallel[100]$) for $a>0$ ($a<0$). 

\begin{figure}[tb]
\includegraphics[width=8.5cm]{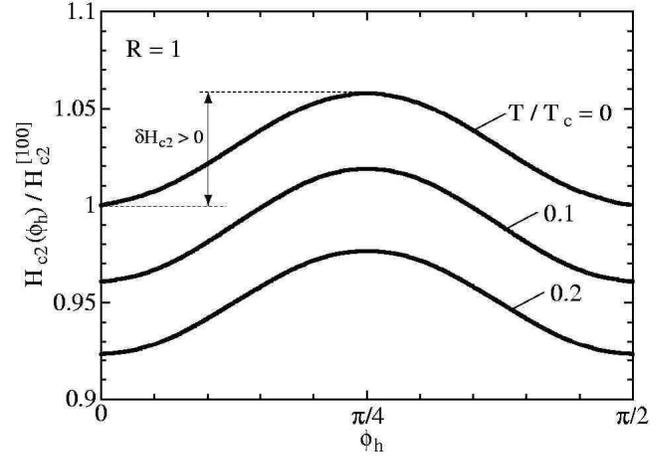}
\caption{The in-plane field angle dependence of the upper critical field with line of zeros in [100] direction ($a=0$, $\phi_0=0$).}
\label{hc2-phi-t}
\end{figure}

\begin{figure}[tb]
\includegraphics[width=8.5cm]{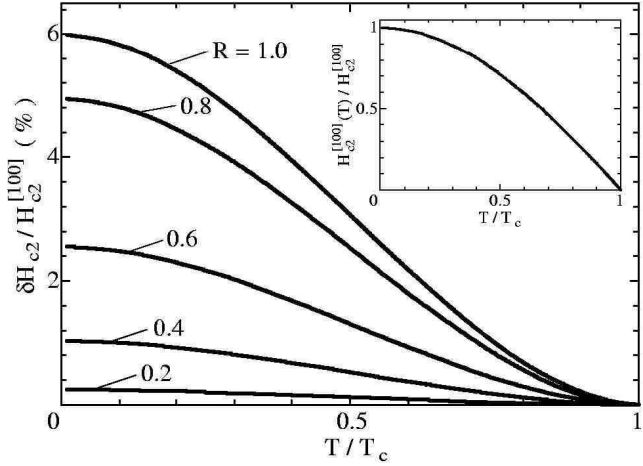}
\caption{The temperature dependence of the oscillatory amplitude of $H_{c2}$ with different size of the gap minimum, $R$ ($a=0$, $\phi_0=0$). The inset shows the temperature dependence of $H_{c2}$ for $H\parallel [100]$.}
\label{hc2-gap}
\end{figure}

At finite temperatures we first discuss the case of the isotropic Fermi velocity, $a=0$.
Figure \ref{hc2-phi-t} shows the in-plane field angle dependence of $H_{c2}$ for $R=1$ at $T/T_c=0$, $0.1$ and $0.2$.
As we expect it shows minima for $\phi_h\parallel\phi_0$ ($=0$ and $\pi/2$).
The oscillatory amplitude in what follows is defined as $\delta A=A(\phi_h\parallel[110])-A(\phi_h\parallel[100])$ where $A$ is either $H_{c2}$ or $C$.
The temperature dependence of $\delta H_{c2}$ scaled by $H_{c2}^{[100]}$ ([100] denotes the field direction.) is shown in Fig.~\ref{hc2-gap}.
The amplitude of the $H_{c2}$ oscillation is a monotonically decreasing function of $T$, and it becomes larger the smaller the size of the gap minimum is.
The inset shows the temperature dependence of $H_{c2}$ for $H\parallel[100]$.

\begin{figure}[tb]
\includegraphics[width=8.5cm]{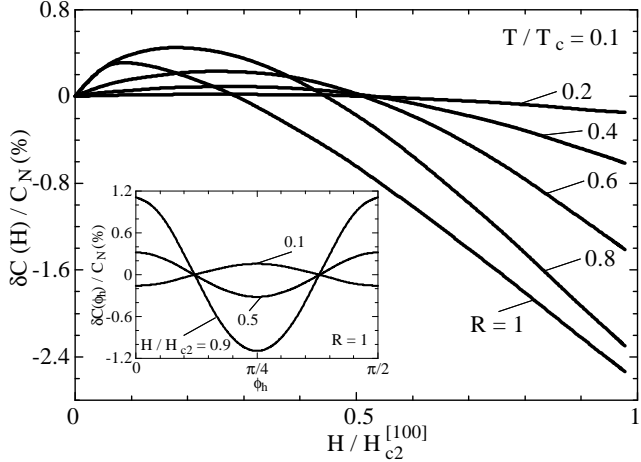}
\caption{The field dependence of the oscillatory amplitude of the specific heat at $T/T_c=0.1$ ($a=0$, $\phi_0=0$). The inset shows the angle dependence of $\delta C$.}
\label{c4-gap}
\end{figure}

The field dependence of $\delta C$ at $T/T_c=0.1$ is shown in Fig.~\ref{c4-gap}.
As the field decreases, the negative amplitude increases and changes sign at $H^*$.
It is emphasized that even for the gap with line of zeros, $R=1$, it reaches the maximum, and again decreases toward to zero.
For larger modulation of the gap $\delta C$ has stronger field dependence.
The overall behavior at the lowest temperature can be understood as follows.
At higher fields the in-plane $H_{c2}$ anisotropy yields the oscillation in $H/H_{c2}$ since $H$ is fixed in the experimental situation.
It leads to the amplitude oscillation in $\Delta$, which predominates the oscillatory behavior in $C$.
Therefore, the oscillatory amplitude of $C$ has the opposite sign of $\delta H_{c2}$.
At lower fields $H\lesssim H^*$, the Doppler shift argument can be applied \cite{Vekhter99}.
The local QP with the momentum $\hat{\mib k}$ has the energy spectrum, $E_{\hat{\mib k}}+{\mib v}_{\rm s}\cdot{\hat{\mib k}}$ in the supercurrent flowing around vortices with the velocity ${\mib v}_{\rm s}$.
This energy shift gives rise to a finite DOS near the gap minima if $|\hat{\mib k}\cdot{\mib v}_{\rm s}|>|\Delta(\hat{\mib k})|$.
Since ${\mib v}_{\rm s}\perp {\mib H}$, the number of activated gap minima becomes largest in the case of $H\parallel \phi_0$ yielding a maximum in $C$.
In the limit of $H\to 0$, $\delta C$ decreases again since the volume fraction of the locally activated QPs becomes small.

\begin{figure}[tb]
\includegraphics[width=8.5cm]{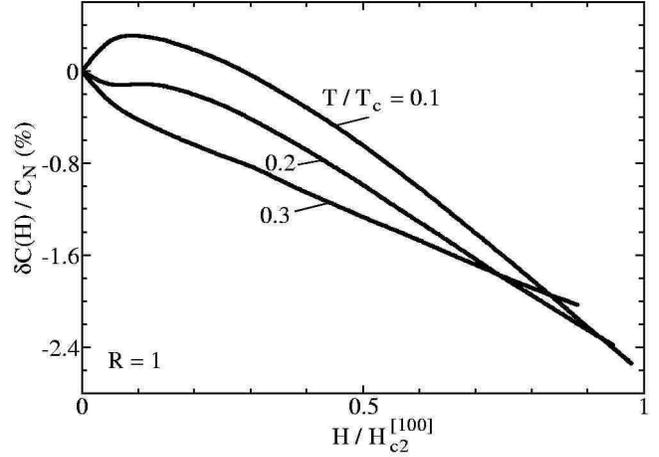}
\caption{The oscillatory amplitude of specific heat at elevated temperature ($a=0$, $\phi_0=0$).}
\label{c4-t}
\end{figure}

It should be noted that the Doppler shift argument is valid only at low temperature.
At elevated temperature, positive contribution to $\delta C$ from the activated QPs near the gap minima becomes less pronounced and the negative contribution from the $H_{c2}$ anisotropy predominates.
This is seen clearly by the field dependence of $\delta C$ at $T/T_c=0.1$, $0.2$ and $0.3$ 
as shown in Fig.~\ref{c4-t}.
For $T/T_c\ge 0.2$ no sign changes occur in $\delta C$.

\begin{figure}[tb]
\includegraphics[width=8.5cm]{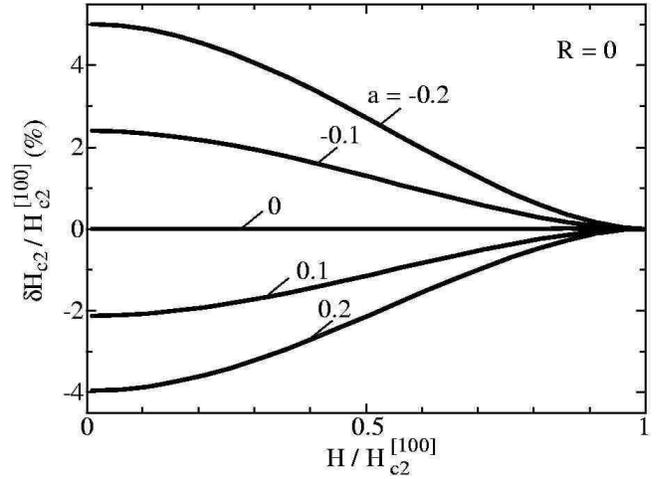}
\caption{The temperature dependence of $\delta H_{c2}$ for the isotropic full gap with the anisotropic Fermi velocity.}
\label{hc2-a}
\end{figure}

Next we discuss the effect of the anisotropy of ${\mib v}$.
For this purpose we consider the isotropic gap, $R=0$.
The temperature dependence of $\delta H_{c2}$ with the anisotropy is shown in Fig.~\ref{hc2-a}.
The effect of anisotropy in $\delta H_{c2}$ has the opposite sign of $a$ and is monotonically decreasing (increasing) function of $T$ for $a<0$ ($a>0$).
Note that the oscillatory amplitude due to the anisotropic velocity can be the same order of magnitude of that due to the gap modulation.

\begin{figure}[tb]
\includegraphics[width=8.5cm]{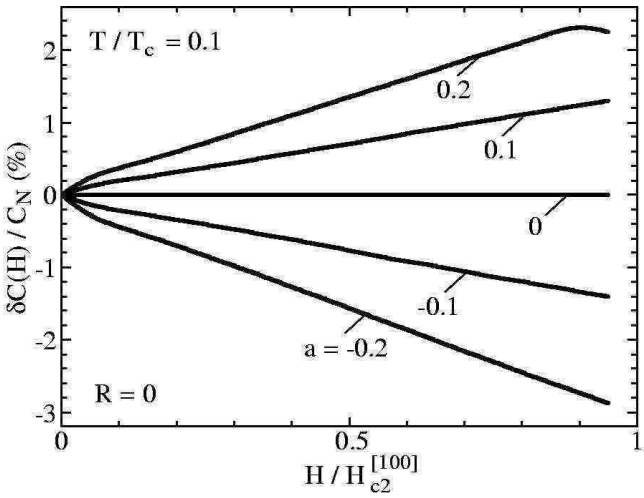}
\caption{The field dependence of $\delta C$ for the isotropic full gap with the anisotropic Fermi velocity.}
\label{c4-a}
\end{figure}

Figure \ref{c4-a} shows the field dependence of $\delta C$ at $T/T_c=0.1$.
Since there is no contribution from the Doppler-shifted QPs in this case ($R=0$), the oscillatory behaviors come purely from the anisotropy of $H_{c2}$, which affects the oscillatory behaviors over weak field region.
Thus, the oscillatory amplitude of $\delta C$ has the same sign of $a$ and its magnitude decreases monotonically as $H$ decreases.

As will be discussed in the next section, the oscillatory behaviors in Sr$_2$RuO$_4$ is determined roughly by the contribution from the primary $\gamma$ band.
Viewing a small positive anisotropy of ${\mib v}$ in the $\gamma$ band \cite{Mazin97}, we expect that the primary gap in the $\gamma$ band has a modulation with an intermediate size of minima in [100] direction ($R\sim$ 0.6--0.8, $\phi_0=0$) for consistency with experiments \cite{Deguchi03,Deguchi04}.

\section{Interplay of two-band superconductivity}
With the knowledge obtained in the previous section, let us discuss the oscillatory behaviors in Sr$_2$RuO$_4$.
Since the $\alpha$ and $\beta$ bands have similar characters of Fermi surface and the Cooper pair scattering between them is expected not to be small, we adopt an effective two-band model for Sr$_2$RuO$_4$.
We use the normal DOS, $n_\gamma=0.57$ and $n_{\alpha\beta}=0.43$, based on the de Haas-van Alphen measurements \cite{Mackenzie03}.
The parameters of two dimensionality are estimated as $\chi_\gamma=6.0\times10^{-6}$ and $\chi_{\alpha\beta}=3.0\times10^{-4}$, respectively, and $\zeta=v_{\alpha\beta\perp}^2/v_{\gamma\perp}^2=4.0$, from the experimental value of $H_{c2}^{(c)}/H_{c2}^{(a)}$ and the Fermi surface topology \cite{Mackenzie03}.
For the gap structure $(R_\ell,\phi_{\ell0})$ and the anisotropy of ${\mib v}$ $(a_\ell)$, we choose $R_\gamma=0.65$, $\phi_{\gamma0}=0$ and $a_\gamma=0.1$ for the primary $\gamma$ band. For the passive $\alpha\beta$ bands, we adopt $R_{\alpha\beta}=1$, $\phi_{\alpha\beta0}=\pi/4$ and $a_{\alpha\beta}=0$, which can explain naturally the anomalous temperature dependence of $\delta H_{c2}$ as will be shown shortly.

\begin{figure}[tb]
\includegraphics[width=8.5cm]{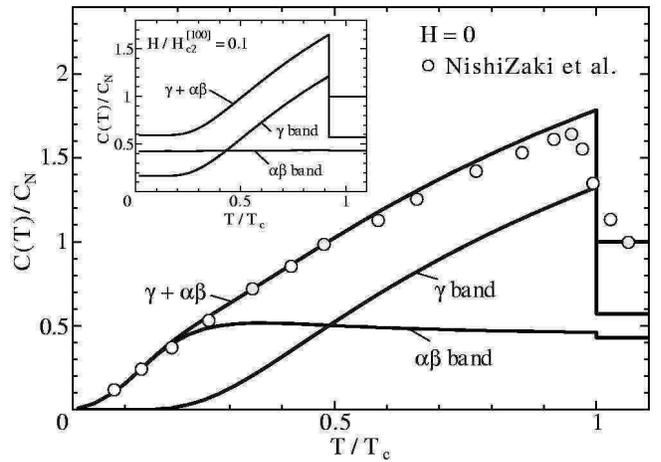}
\caption{The temperature dependence of $C$ at zero field. The open circle represents the experimental data taken from Ref.~\cite{NishiZaki00}. The inset shows $C(T)$ at $H/H_{c2}^{[100]}=0.1$.}
\label{c-t-2}
\end{figure}

Before discussing the oscillatory behaviors we first determine the interaction parameters for Sr$_2$RuO$_4$.
This is done by the specific heat fitting at $H=0$.
The best fitting is achieved by
\begin{equation}
\hat{\lambda}=\left(\begin{array}{cc}\lambda_1 & \lambda \\ \lambda & \lambda_2\end{array}\right),
\end{equation}
with $\lambda_1=0.4$, $\lambda_2/\lambda_1=0.65$ and $\lambda/\lambda_1=0.12$, as shown in Fig.~\ref{c-t-2}.
The open circle represents the experimental data taken from Ref.~\cite{NishiZaki00}.
The inset shows the results at $H/H_{c2}^{[100]}=0.1$, which indicate that almost all the QPs in the $\alpha\beta$ band is already activated at this field.

\begin{figure}[tb]
\includegraphics[width=8.5cm]{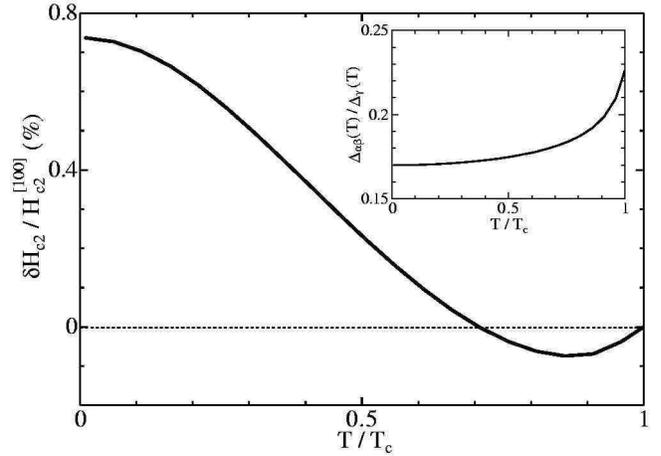}
\caption{The field dependence of $\delta H_{c2}$. The ratio of the two gaps at $H_{c2}(T)$ is shown in the inset.}
\label{h4-t-2}
\end{figure}

Figure \ref{h4-t-2} shows the temperature dependence of the in-plane $H_{c2}$ anisotropy.
The behavior at low temperatures is similar to the single $\gamma$ band case in Fig.~\ref{hc2-gap} except that the oscillatory amplitude is suppressed by the positive weak anisotropy in ${\mib v}$.
It is remarkable that the oscillatory amplitude changes its sign near $T_c$.
This can be understood as an interplay of the primary gap in $\gamma$ band having the minimum in [100] direction with the secondary gap in $\alpha\beta$ band having the different minimum in [110] direction.
The contribution from the secondary gap, which has the opposite sign of the primary gap's, is enhanced near $T_c$ yielding the negative amplitude in $\delta H_{c2}$.
The enhancement of the ratio of two gaps at $H_{c2}(T)$ is shown in the inset of Fig.~\ref{h4-t-2}.
This is the strong evidence that the secondary gaps in $\alpha\beta$ bands have the gap minima in [110] direction.
The similar interplay of two-band superconductivity is observed in MgB$_2$, in which the 3-dimensional passive $\pi$ band predominates over the 2-dimensional primary $\sigma$ band showing the strong temperature dependence in $H_{c2}^{(a)}/H_{c2}^{(c)}$ \cite{Lyard02,Angst02,Dahm03,Zhitomirsky04,Miranovic03a}.

\begin{figure}[tb]
\includegraphics[width=8.5cm]{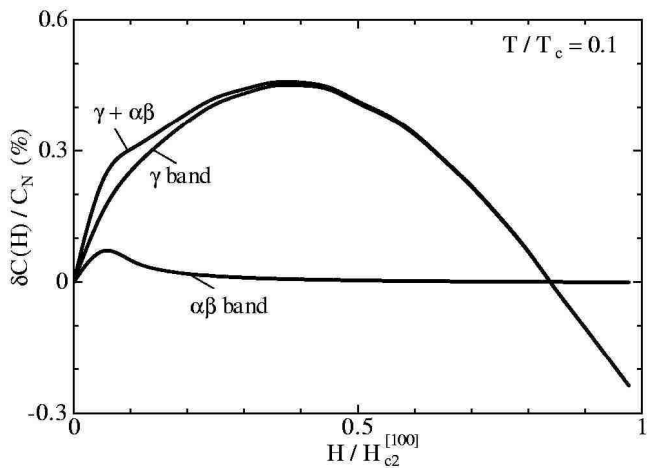}
\caption{The field dependence of $\delta C$.}
\label{c4-h-2}
\end{figure}

Finally, we discuss the field dependence of $\delta C$ for Sr$_2$RuO$_4$.
The contributions from each bands are separately shown in Fig.~\ref{c4-h-2}.
Both the gap modulation with minima in [100] direction and the anisotropy in ${\mib v}$ with $a>0$ contributes to the positive oscillatory amplitude in $\delta C_\gamma$ at low fields.
Since the passive $\alpha\beta$ bands give no contribution to $\delta C$ except in very low fields, total amplitude becomes smaller than the single-band result by a factor of $n_\gamma$(=0.57).
It is remarkable that the $\alpha\beta$ contribution at low fields has positive sign.
We expect naively that the gap modulation with minima in [110] direction gives the negative sign of $\delta C_{\alpha\beta}$ at low fields.
However, this naive expectation based on the Doppler shift argument is valid only at very low temperature.
Since the passive gap is small as compared with $\Delta_\gamma(T)$ even at low fields, $T/T_c=0.1$ is regarded effectively in high temperature region.
Therefore, the positive contribution from $\alpha\beta$ bands results as the case of $T/T_c\ge 0.2$ shown in Fig.~\ref{c4-t} (note that $\phi_0=\pi/4$ in this case).
The overall behaviors of $\delta C$ is consistent with experiments \cite{Deguchi03,Deguchi04}.

\section{Summary}
We have investigated the influence of the modulation of the gap and the anisotropy of the Fermi velocity to the oscillatory behaviors of $H_{c2}$ and $C$, using the approximate analytic solution of the quasiclassical formalism.

It is shown that for the single-band model the minimum direction of the in-plane $H_{c2}$ anisotropy coincides with either the minimum of the gap modulation or the Fermi velocity.
If they are in different directions, the resultant amplitude becomes small due to the competing contributions from both effects.
The absolute magnitude of the $H_{c2}$ anisotropy is a monotonically decreasing function of $T$.
Due to the $H_{c2}$ anisotropy the oscillatory behavior of $C$ at high fields shows the opposite tendency of the $H_{c2}$ anisotropy and its magnitude increases toward $H_{c2}$.
At low fields and low temperatures the effect of the so-called Doppler shift plays an important role yielding the opposite oscillatory behaviors to the high-field ones.
Therefore we can safely conclude that the source of the oscillatory behaviors is the modulation of the gap, if a sign change is observed in the oscillatory amplitude of $C(H)$.
It is noted however that the effect of the Doppler shift is pronounced only at sufficiently low temperatures.

With a combination of the results above, we have discussed the gap symmetry for Sr$_2$RuO$_4$.
The oscillatory behavior of $C$ is roughly determined by the primary gap in the $\gamma$ band.
To be consistent with experiments we conclude that the primary gap has an intermediate magnitude of minima in [100] direction.
Assuming that the secondary gap in $\alpha\beta$ bands has line of zeros in [110] direction we  have shown that the in-plane $H_{c2}$ anisotropy exhibits the anomalous temperature dependence with a sign change near $T_c$.
This is understood by the opposite contribution from the passive bands, which becomes predominant over the $\gamma$ band contribution near $T_c$.
The observation that the interplay of two gaps is pronounced near $T_c$ provides an alternative way to determine gap symmetry in passive bands, which is hardly accessible by ordinary experiments.
With the assumed gap model the total behavior of $\delta C$ together with the $\alpha\beta$ band contributions is consistent with experiments.

\section*{Acknowledgment}
The author would like to thank T.M. Rice, M. Sigrist, I. Mazin, Y. Matsuda, K. Izawa, T. Watanabe, N. Nakai, K. Deguchi, T. Nomura, K. Kubo and Y. Kato for fruitful discussions.

\end{document}